# Multi-functional volumetric meta-optics for color and polarization image sensors

PHILIP CAMAYD-MUÑOZ, CONNER BALLEW, GREGORY ROBERTS, ANDREI FARAON*

*Kavli Nanoscience Institute and Thomas J. Watson Sr. Laboratory of Applied Physics, California Institute of Technology, Pasadena, California 91125*
*Corresponding author: faraon@caltech.edu*

**Three-dimensional elements, with refractive index distribution structured at sub-wavelength scale, provide an expansive optical design space that can be harnessed for demonstrating multi-functional free-space optical devices. Here we present 3D dielectric elements, designed to be placed on top of the pixels of image sensors, that sort and focus light based on its color and polarization with efficiency significantly surpassing 2D absorptive and diffractive filters. The devices are designed via iterative gradient-based optimization to account for multiple target functions while ensuring compatibility with existing nanofabrication processes, and experimentally validated using a scaled device that operates at microwave frequencies. This approach combines arbitrary functions into a single compact element even where there is no known equivalent in bulk optics, enabling novel integrated photonic applications.**

Historically optical design has been modular, where different optical elements are combined to achieve complex functionality. This paradigm provides an intuitive way to build and reconfigure optical setups. Recent advancements in nano-fabrication technologies have enabled the synthesis of multiple functions into a single optical element with sub-wavelength features [1]. Examples include metasurface lenses that can split different polarizations and spectral bands [2–5]. However, the performance and multi-functionality that can be achieved with metasurfaces and other planar structures is inherently limited by the number of optical modes that can be controlled, which scales with the volume of the device and the maximum refractive index contrast [6,7].

The ultimate optical design space is a three-dimensional volume where the index of refraction can be specified with sub-wavelength spatial resolution, which maximizes the available degrees of freedom. However, it remains a challenging inverse design problem to identify the 3D index distribution that will perform a desired optical function: in the high-contrast limit, the optical scattering function is dominated by multiple scattering events and cannot be easily inverted via direct Fourier-based methods [8]. Instead we apply topology optimization [9–13] to design high-efficiency, strongly scattering devices: we efficiently compute the sensitivity of device performance with respect to perturbations of the refractive index distribution using full-wave simulations and the adjoint method. The design is then updated via gradient descent, allowing us to iteratively generate designs that achieve the target scattering function while conforming to material and fabrication constraints. This approach can combine arbitrary functions into a single compact element even where there is no known equivalent in bulk optics, and has been applied to a range of integrated photonic applications [14–16].

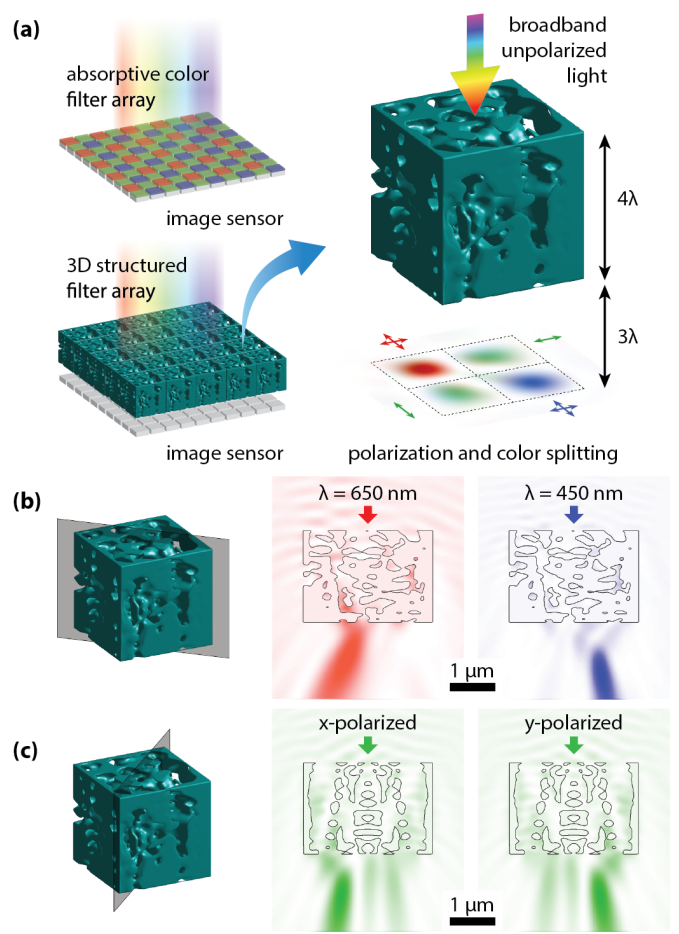

**Fig. 1.** Inverse-designed color and polarization sorter. (a) Spectral filter consisting of a 3D nanostructured polymer cube. Incident light is efficiently focused into four sub-pixels on a focal plane below the device. (b) Red (600 nm – 700 nm) and blue (400 nm – 500 nm) spectral bands are sorted into opposite quadrants. The color indicates field intensity. The brightness of the output field intensity demonstrates the focusing nature of the devices. (c) Similarly, the green (500 nm – 600 nm) band is further split according to linear polarization. Red and blue quadrants are polarization-independent.

In order to demonstrate the versatility of high-contrast volumetric structures, we present a 3D dielectric element with lateral dimensions of a few wavelengths, which focuses incident light to different locations based on frequency and polarization. This element can be deployed as a color filter array on image sensors (Fig. 1a), improving low-light sensitivity while maintaining high color contrast [17,18]. Focusing is mediated by multiple-scattering within the volume, resulting in high color and polarization selectivity at the focal plane. We explore two different material configurations, one suitable for high resolution 3D printing [19] and another for large-scale multilayer lithography [20]. Finally, we experimentally validate a design scaled to operate at microwave frequencies.

The scattering element is designed using the adjoint optimization method [9–13], which iteratively generates a structure to optimize a specified objective function (See Methods and Supplementary Sections I and II). Here an incident plane wave is simultaneously focused at different positions for multiple frequencies. Through repeated iterations the design converges to a complex three-dimensional topology (Fig. 1a). This freeform design can be fabricated via direct write lithography, which is capable of producing arbitrary 3D structures with sub-wavelength spatial resolution even at visible wavelengths [21,22]. The spectral filter is designed within a $(2~\mu m)^3$ cube composed of polymer [23] (index = 1.56) and air. The polymer index choice is not necessarily optimal for optical performance, but is instead motivated by the availability of specific materials for direct write lithography. The $(2~um)^3$ device volume is discretized into a $100\times100\times100$ grid, where each $(20~nm)^3$ voxel is assigned to either polymer or air. A projection filter [24] is used to restrict the optimization to binary index structures without imposing constraints on the minimum feature size. However, the resulting device contains no features smaller than three voxels wide, for a minimum feature size of 60 nm. When illuminated from above, the optimized scattering element efficiently focuses red (600 nm – 700 nm), green (500 nm – 600 nm) and blue (400 nm – 500 nm) visible light onto one of four target regions (each 1 μm × 1 μm) at a focal distance of 1.5 μm below the cube (Fig. 1).

This design method provides enormous flexibility in defining the target scattering function, with independent control for any incident polarization, angle, or frequency. Therefore, we can incorporate additional functionality beyond standard color filters. Here, we repurpose the redundant green pixel of the 2×2 Bayer pattern in order to discriminate between two orthogonal linear polarizations, assigning each polarization to opposing quadrants. Fig. 1b,c shows the simulated intensity of incident light within the cube for different wavelengths, plotted along a diagonal cross section that intersects the red and blue quadrants. Each wavelength undergoes multiple scattering within the volume of the cube before converging at its respective target region. The device sorts normal incident red, green, and blue light with 84%, 60% and 87% efficiency, respectively, where efficiency is defined as the fraction of the total power incident on the cube that reaches the target quadrant averaged across the target spectral band (See Supplementary Section III).

Large-scale implementation of these devices in image sensors will require high-throughput fabrication with sub-100-nm resolution. This can be achieved by multi-layer lithography, where three-dimensional devices are constructed through repeated deposition and patterning of two-dimensional layers [20]. We design a multi-layer spectral filter by directly incorporating fabrication constraints into the design algorithm. Fig. 2a shows a layered design composed of $TiO_2$ and $SiO_2$, materials that are transparent at visible frequencies. The constrained device consists of five 2μm × 2μm layers, each 400 nm tall. Each layer contains a set of irregular $TiO_2$ mesas surrounded by $SiO_2$ (Fig. 2b) and the minimum feature size is limited to 60 nm (Supplementary Section II). These dimensions are achievable in state of the art cleanrooms. Each $(2~um)^2$ layer is discretized into a 100x100 grid and the combination of a blurring filter used during design and a postprocessing step ensure the desired minimum feature size of 60nm.

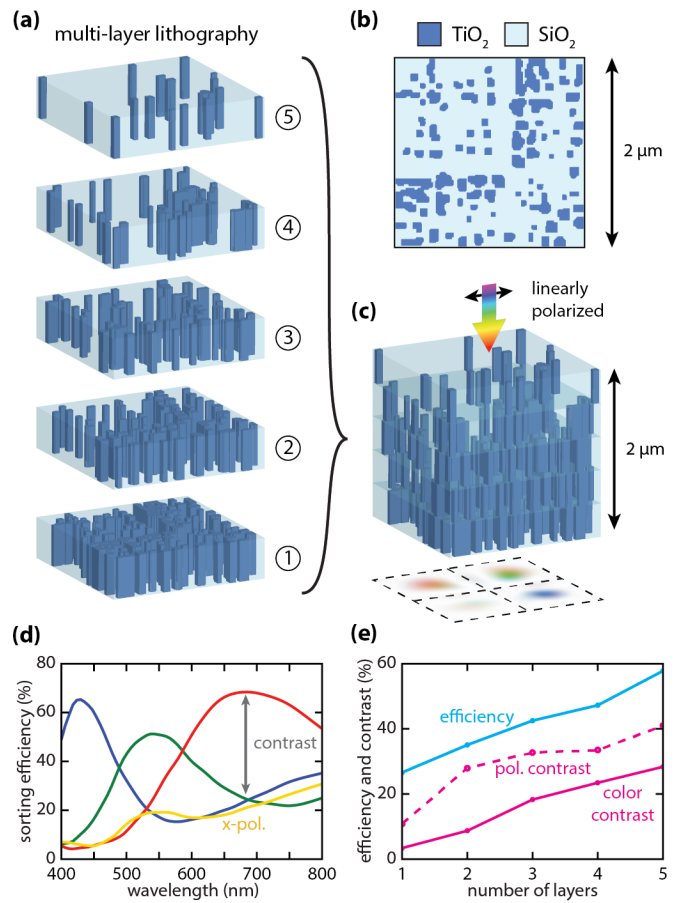

**Fig. 2.** Spectral filter designed for multi-layer lithography. (a) Three-dimensional scattering element constructed by stacking multiple two-dimensional layers. (b) Each layer consists of a series of $TiO_2$ mesas, which can be fabricated using standard lithography and material deposition. Posts are infiltrated with a $SiO_2$ matrix, forming a flat substrate for subsequent layers. (c) When combined, the five-layer stack performs the same function as the freeform design shown in Fig. 1. Linearly polarized light is focused into distinct sensor regions depending on frequency and polarization. (d) Sorting efficiency within the visible spectrum, defined as the fraction of incident power reaching the target quadrant as indicated in (c). The yellow curve indicates light present at the cross-polarized quadrant. See Supplementary for movies showing focal plane intensity as the input wavelength is swept across the design bandwidth. (e) Spectrum-averaged sorting efficiency, polarization contrast, and color contrast for devices with increasing number of layers, each 400-nm-thick.

The combined layers perform the same function as the freeform design, sorting visible light according to frequency and polarization. Fig. 2c shows the simulated total intensity at the focal plane when the filter is illuminated from above by broadband, linearly polarized light. The color corresponds to the hue of the light, and demonstrates efficient color sorting into the target quadrants. The lower-left quadrant is assigned to the orthogonal polarization of incident light, and therefore high power transmission. Due to the diagonal mirror symmetry of the design, the device is equally efficient at sorting an orthogonal polarization, adding polarization selectivity in the green pixels. Although this device is designed to sort light at normal incidence, the functionality is maintained at oblique incidence angles up to ±8°. See Supplementary VI for details.

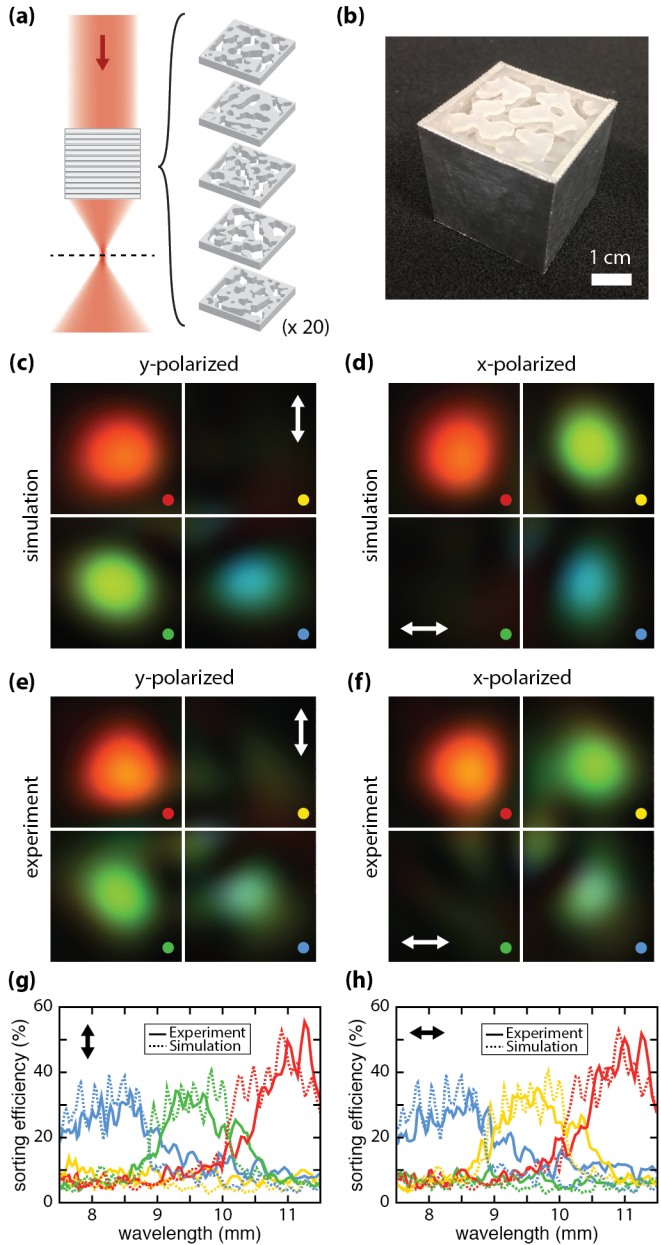

We investigate the trade-off between multi-functionality and device thickness by designing a series of devices with different numbers of layers. Each device follows the same design algorithm using 400 nm layers. Fig. 2e shows the sorting efficiency of each device, averaged across the visible spectrum. While the single-layer metasurface performs marginally better than empty space, efficiencies grow steadily with device thickness (for details see Supplementary Section IV and Fig. S4). In addition, thicker devices exhibit improved color and polarization contrast. Contrast is defined here as the difference in normalized power between the two brightest quadrants, and therefore reflects the capacity to distinguish incident colors and polarizations. With five layers, the volumetric scattering element outperforms an absorptive filter with respect to sorting efficiency (57%), color contrast (29%), and polarization contrast (41%).

The chosen dimensions can be fabricated in a semiconductor cleanroom with significant investment in process development, but they are challenging to achieve in a university cleanroom with limited resources. Thus, we experimentally validate the fabrication-constrained design using a centimeter-scale analog (Fig. 3) operating at microwave frequencies in the $K_a$ band (26–40 GHz). This device is designed to maximize the sorting efficiency at 18 wavelengths, equally spaced across the $K_a$ band, and for both orthogonal linear polarizations. The device is constructed from a stack of 20 polypropylene (*PP*) sheets (index = 1.5), each 1.6-mm-thick and patterned with 1mm minimum feature size. A metallic boundary on the sides of the cube is incorporated to limit interference from the measurement apparatus. The microwave device occupies a 35mm × 35mm footprint, the same as its optical analog relative to the operating wavelength. The index contrast of the scattering element is also similar in both cases: $n_{TiO_2}/n_{SiO_2}$ = 1.67, $n_{PP}/n_{air}$ = 1.5. Fig. 3b shows the assembled device.

We characterize the performance of the spectral filter by measuring the complex microwave field scattered by the device when illuminated by a collimated Gaussian beam. Fig. 3 c–f shows the simulated and measured intensity of the microwave fields at the focal plane of the device, summed over all frequencies within the measurement bandwidth. The color corresponds to the observed hue of the analogous optical fields when scaled to visible frequencies. These field profiles show efficient color and polarization sorting into each target quadrant, with good agreement between experiment and simulation. Fig. 3 (g) and (h) show the relative sorting efficiency of the spectral filter across the measurement spectrum. These efficiencies are defined as power transmitted through each target quadrant, normalized to the total power at the focal plane. We observe close agreement between experimental and simulated efficiencies, shown as solid and dashed curves, respectively. Each band shows efficient sorting with low crosstalk from out-of-band light, roughly 10%. The individual peaks in within each band correspond to the optimized frequencies. Sharp transitions between spectral bands highlight the improved color discrimination over typical dispersive scattering elements. We observe similar discrimination between orthogonal polarizations, shown in green and yellow curves. The remaining intensity is lost to reflection and oblique scattering outside of the target area. More details on the design and measurement of the microwave device are provided in Methods and Supplementary Sections VI and VII.

**Fig. 3.** Microwave measurement and simulation results. (a,b) Microwave filter composed of patterned polypropylene sheets, assembled into a cube. The cube is illuminated by a collimated microwave source operating in the $K_a$ band (26 – 40 GHz). (c,d) Field intensity at the focal plane of the microwave filter for linearly polarized illumination, as obtained by full wave simulation. Colors represent the equivalent hue of visible light, scaling the microwave frequency by a factor of ×17,500. (c) and (d) correspond to vertical and horizontal polarization, respectively. (e,f) Measured microwave field intensity at the focal plane. See Supplementary for movies showing focal plane intensity as input wavelength is swept across the measurement bandwidth. (g,h) Measured and simulated microwave sorting efficiency for each polarization. Dashed lines represent simulated efficiency, while solid lines represent experimental values. Each color corresponds to a quadrant as indicated by the dots in (c–f).

remains dark. We characterize the sorting efficiency of the filter across the visible spectrum (Fig. 2d), demonstrating both color contrast and

We have demonstrated a dielectric scattering element that efficiently sorts light according to its polarization and frequency. These devices can be tiled in a two-dimensional filter array, providing color selectivity for image sensors without sacrificing transmitted power. Although the design is inherently complex, drawn from a high-dimensional design space, we explicitly impose realistic design constraints to ensure that these devices could be fabricated at large scale. The design does not rely on the local effective-medium assumptions or high index contrast. In particular, these filters can be fabricated using multi-layer fabrication with achievable requirements on feature size and number of layers.

We improve upon existing dielectric filters by encoding the sorting functionality into the complex multiple-scattering within a volume, rather than at a single surface. In effect, the scattering function is decorrelated for each frequency, polarization, and spatial mode [25]. Coupled with gradient-based inverse design, this approach provides enormous freedom to define arbitrary scattering elements with tailored functionality. This design space encompasses recent demonstrations of cascaded diffractive elements [26], while including effects of strong multiple scattering. Here, we add polarization selectivity to the standard color filter for expanded sensing modalities [27]. The volumetric scattering element can support both narrowband and broadband resonances, including discontinuous spectra, without relying on chromatic dispersion. Beyond imaging, we may tailor each pixel to collect a spectrum of interest, such as chemical fluorescence, for use in remote sensing applications [28]. Finally, this approach could be used to control the scattering based on the spatial mode of incident light for high-NA imaging [29], angular selectivity in photovoltaics [30], or automatic object recognition [31].

**Funding.** This work is supported by the DARPA EXTREME program, grant HR00111720035, and NIH 1R21EY029460-01

**Acknowledgment.** The authors thank Maria Alonso-delPino for assistance and advice regarding near-field microwave imaging. The measurements were carried out at the Jet Propulsion Laboratory, California Institute of Technology, under contract with the National Aeronautics and Space Administration.

**Disclosures.** The authors declare no conflicts of interest

See Supplement 1 for supporting content.

# Multi-functional volumetric meta-optics for color and polarization image sensors: supplementary material

PHILIP CAMAYD-MUÑOZ, CONNER BALLEW, GREGORY ROBERTS, ANDREI FARAON*

*Kavli Nanoscience Institute and Thomas J. Watson Sr. Laboratory of Applied Physics, California Institute of Technology, Pasadena, California 91125*
*Corresponding author: faraon@caltech.edu*

## I. Inverse design method

In this work we design three-dimensional dielectric structures, optimized to perform a specified optical scattering function: in this case, focusing incident plane waves to different positions depending on the frequency and polarization. The structure is defined by a spatially-dependent refractive index distribution $n(\vec{x})$ within a cubic design region. This represents an expansive design space with the capacity to express a broad range of complex optical multi-functionality. However, identifying the optimal index distribution for a given target function remains a challenging inverse design problem, particularly for strongly scattering devices [1]. Therefore, we adopt an iterative approach guided by gradient descent (Fig. S1) in order to efficiently generate complex 3D designs [2]: starting from an initial index distribution, we use full-wave simulations (FDTD) to calculate the sensitivity of the sorting efficiency with respect to perturbations of the refractive index. The sensitivity can be calculated from just two simulations, allowing efficient optimization of three-dimensional devices with modest resources. Based on the sensitivity, we modify the initial design in order to maximize the performance while conforming to fabrication constraints. This update process is repeated until the optimized device can efficiently perform the target function.

For the designs presented in this manuscript, we start with a uniform refractive index distribution, $n_0(\vec{x}) = \frac{1}{2}(n_{max} + n_{min})$. This distribution is continually updated to maximize the electromagnetic intensity at the target location in focal plane, $f(n(\vec{x})) = |\vec{E}(\vec{x}_0)|^2$. This figure of merit serves as a proxy for sorting efficiency while simplifying the sensitivity calculation. The sensitivity, $\frac{df}{dn}(\vec{x})$, is computed from the electromagnetic fields in two FDTD simulations:

$$\frac{df}{dn}(\vec{x}) = 2n(\vec{x})\operatorname{Re}\left\{\vec{E}_{fwd} \cdot \vec{E}_{adj}\right\} \quad \text{(S1)}$$

where $\vec{E}_{fwd}$ are the electric fields within the cube when illuminated from above with a plane wave, and $\vec{E}_{adj}$ are the electric fields within the cube when illuminated from below with a point source at the target location. The phase and amplitude of the point source are given by the electric field at the target location in the forward simulation. We simultaneously calculate the sensitivity for multiple incident wavelengths and polarizations across the visible spectrum, assigning each spectral band to a different quadrant: red (600 nm – 700 nm) green (500 nm – 600 nm) and blue (400 nm – 500 nm). Then we use the spectrally-averaged sensitivity to update the refractive index of the device:

$$n_{i+1}(\vec{x}) = n_i(\vec{x}) + \alpha \sum_\lambda \frac{df_\lambda}{dn}(\vec{x}) \quad \text{(S2)}$$

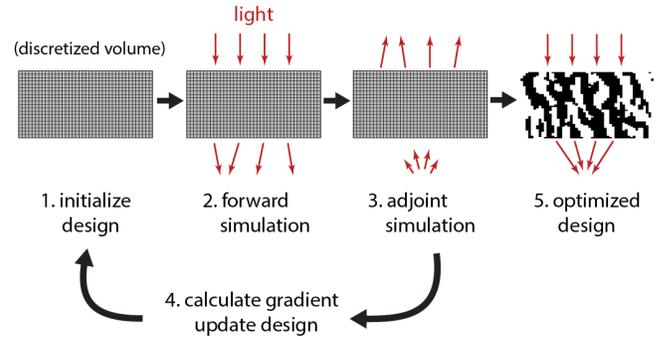

**Fig. S1.** Scattering elements are tailored to perform a target function through gradient descent. Starting from an initial design, we compute the sensitivity of the sorting efficiency to changes in the refractive index using a pair of FDTD simulations. The design is then updated in proportion to the sensitivity. After many iterations, the resulting structure focuses incident light with high efficiency.

The step size $\alpha$ fixed at a small fraction (typically $\alpha = 0.001$) to ensure that the change in refractive index can be treated as a perturbation in the linear regime. The sensitivity is recalculated after each update.

## II. Fabrication constraints

### A. Binary index

During the optimization process, we directly enforce a set of constraints on the index distribution as required by the fabrication process. In particular, we are designing high-contrast scattering elements constructed from two materials. Although the gradient descent algorithm detailed above produces optimized devices with gradient index, we can enforce the binary condition by introducing an auxiliary density $\rho(\vec{x})$ ranging from [0,1]. The density is related to the refractive index distribution via a sigmoidal projection filter [3]:

$$n(\vec{x}) = P(\rho(\vec{x})) = \left(\frac{1}{2} + \frac{\tanh(2\beta\rho(\vec{x}) - \beta)}{2\tanh(\beta)}\right)(n_{max} - n_{min}) + n_{min} \quad \text{(S3)}$$

where the parameter $\beta$ controls the filter strength. For small $\beta$, the index distribution is equal to the density scaled to the range of available refractive index. For large $\beta$, the sigmoid filter approximates a Heaviside function, and the index distribution pushed toward either extreme. Importantly, the filter function is continuously differentiable, such that the sensitivity can be written in terms of the density: $\frac{df}{d\rho} = \frac{df}{dn}\frac{dn}{d\rho}$. During optimization we can parameterize the design using the density $\rho(\vec{x})$ and $\beta$, gradually increasing the strength of the filter. At early stages and small $\beta$, this is equivalent to the unfiltered case. Over time as the strength increases, the optimized index distribution is gradually pushed toward a binary design, even as the density remains continuous.

### B. Minimum feature size

In addition to material constraints, device designs must conform to the resolution limits imposed by the fabrication process. For example, diffraction and proximity dosing effects limit electron beam lithography to approximately 10 nm features. We enforce this minimum feature size for device designs by introducing a "dilated" density $\tilde{\rho}(\vec{x})$, which represents the maximum density $\rho(\vec{x}')$ within a neighborhood $\Omega$ of each point $\vec{x}$ [4].

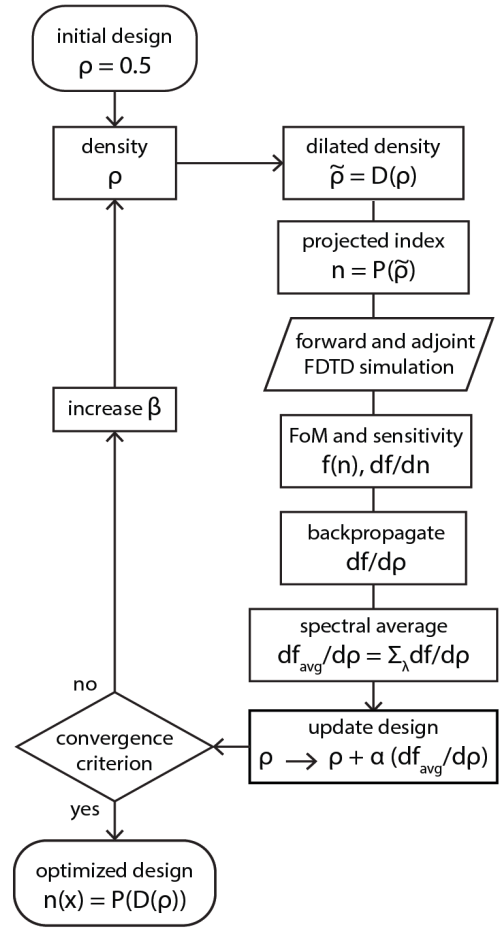

**Fig. S2.** The design of optimized scattering elements consists of iterative updates to the refractive index distribution, guided by the sensitivity to perturbations in the index, and constrained by materials and fabrication processes as outlined above.

$$\tilde{\rho}(\vec{x}) = D(\rho(\vec{x})) = \sqrt[M]{\frac{1}{M}\sum_\Omega (\rho(\vec{x}'))^M} \quad \text{(S4)}$$

For a sufficiently large exponent $M$, this operation approximates morphological grayscale dilation. However it is continuously differentiable with respect to the arguments. Therefore, the sensitivity can be written in terms of the un-dilated density: $\frac{df}{d\rho(\vec{x})} = \frac{df}{d\tilde{\rho}(\vec{x})} = \sum_\Omega \frac{df}{d\tilde{\rho}(\vec{x}')}\frac{d\tilde{\rho}(\vec{x}')}{d\rho(\vec{x})}$. The neighborhood $\Omega$ is taken to be a circle, where the radius represents the minimum feature size.

When combined with the projection filter, the dilation operation eliminates small features in the final binary index distribution. During optimization, the device is parameterized by the density $\rho(\vec{x})$, while the index is defined by the dilated density $\tilde{\rho}(\vec{x})$:

$$n(\vec{x}) = P(\tilde{\rho}(\vec{x})) = P(D(\rho(\vec{x}))) \quad \text{(S5)}$$

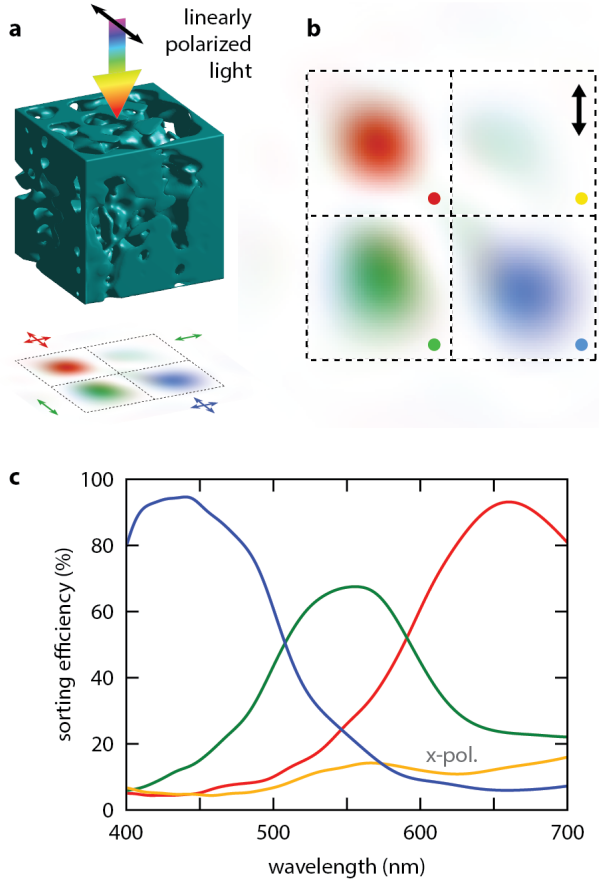

**Fig. S3.** Sorting efficiency of 3D printed scattering element. (a) Device consists of a single piece of polymer, as shown in Fig. 1. Light incident from above is focused to different quadrants depending on the polarization and frequency. (b) Intensity distribution at the focal plane under broadband, linearly polarized illumination. Colors correspond to the observed hue for visible light. The upper right quadrant is assigned to cross-polarized illumination, and therefore remains dark. (c) Sorting efficiency for each of the four focal regions, defined as the fraction of incident light focused into the target quadrant.

The neighborhood $\Omega$ is taken to be a circle, where the radius represents the minimum feature size.

*C. Connected layered designs*

Some of the device designs are intended for fabrication by multi-layer 2D lithography, consisting of several patterned slabs that are invariant in the vertical direction (Fig. 2,3). In this case, we restrict the optimization by averaging the calculated sensitivity in the vertical direction within each layer. In effect, voxels within each layer are governed by a shared 2D profile.

For the microwave device (Fig. 3), the design is further constrained so that each layer is fully connected with no floating pieces. We directly impose connectivity by periodically adding bridges between disconnected islands within each layer. This intervention does not take sensitivity into account, and typically causes a small decrease in device performance. Therefore, we only apply the connectivity constraint once per 40 iterations, allowing the performance to recover thereafter.

## III. Sorting efficiency of 3D printed design

The freeform scattering element shown in Fig. 1 is designed for high-resolution 3D printing. It consists of a monolithic polymer cube (index = 1.56), infiltrated by a series of holes (Fig. S3a). As with all the designs presented in this manuscript, the device sorts incident light to different locations depending on the wavelength and polarization. Here we show the sorting efficiency under linearly polarized illumination. The visible spectrum is divided into three spectral bands (400 nm – 500 nm, 500 nm – 600 nm, 600 nm – 700 nm), each focused on a different quadrant in the focal plane (Fig. S3b). We impose mirror symmetry along a diagonal plane bisecting the red and blue quadrants. This guarantees that the green spectral band will focus orthogonal polarizations to opposite quadrants, while the red and blue spectral bands will be polarization independent. Fig. S3c shows the sorting efficiency for each quadrant across the visible spectrum, defined as the fraction of incident optical power that reaches the target quadrant. This design achieves greater efficiency than similar layered devices (Fig. 2d), which impose stronger restrictions on the device topology.

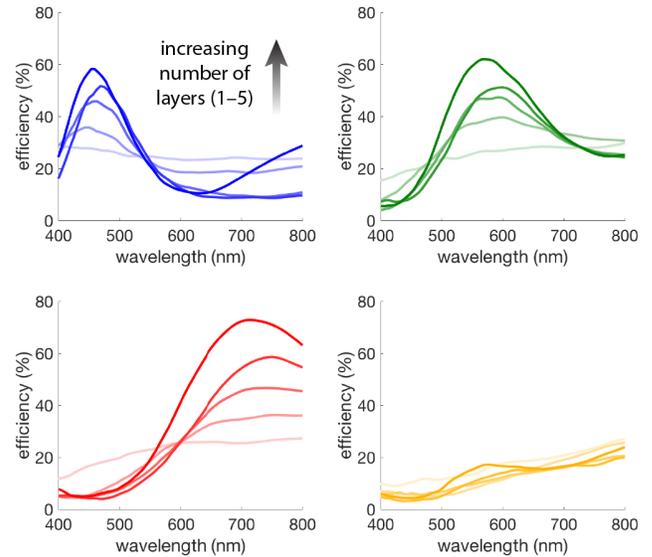

**Fig. S4.** Sorting efficiency for increasing device thickness. Efficiency spectra for the multilayer designs shown in Fig 2, optimized for 1–5 device layers. Each plot corresponds to a different focal region, including the cross-polarized sub-pixel shown in yellow. Darker lines indicate increasing device thickness.

## IV. Dependence on device thickness

Efficient multifunctional devices require a large number of optical degrees of freedom, which scales with the overall device thickness. In the case of spectral- and polarization sorting elements, this restricts the performance of ultrathin devices. Here we show how complex sorting behavior emerges with increased device thickness. Fig. S4 shows the sorting efficiency of several layered devices (similar to Fig. 2), optimized with overall thicknesses ranging from 400 nm – 2000 nm. Thin devices are unable to effectively sort incident light. As thickness increases, the efficiency improves while crosstalk is minimized. We believe that adding even more layers than shown here will increase the performance of the optimized device although not indefinitely. The effect of the number of layers used as well as the thickness of the individual layers is an interesting topic for future investigation.

## V. Dependence on incident angle

The devices presented in this manuscript were designed to sort light at normal incidence. In realistic imaging systems, the illumination will span a range of angles that depends on the numerical aperture of the imaging optics. For example, a typical smartphone camera [5] has a numerical aperture of NA = 0.21, corresponding to an acceptance cone spanning ±7.8° in $SiO_2$. In order to evaluate the device performance under realistic conditions, we have characterized the sorting efficiency of the layered design (shown in Fig. 2) for different illumination angles. As the angle deviates from the surface normal the efficiency decreases (Fig. S5). This is due to an increase in scattering into non-targeted quadrants within the focal plane, as well as reflection from the surface. The drop-off in performance with incident angle is a consequence of decorrelation in the angular scattering through disordered media [6], which scales with the thickness of the scattering medium. Since the device maintains high performance near normal incidence, we define the steepest functional angle where the performance drops to half of the maximum efficiency, roughly ±8°. Therefore, this design may be deployed in color and polarization filters in mobile imaging systems. The range of incidence angles can be extended to steeper angles by reducing the device thickness. Furthermore, we may directly include non-normal incidence into the design algorithm [2], which has so far only considered normal incidence.

## VI. Microwave near field measurement

We sample the scattered microwave near field in the $K_a$ band using a scanning antenna. An incident Gaussian beam (FWHM = 25 mm) is generated by a vector network analyzer coupled to free space via a microwave horn antenna and focusing mirror (Fig. S6). The input beam passes through the cube, scattering into the far field. We sample the local electric field at a measurement plane 62 mm beyond the output aperture of the device using a WR-28 waveguide flange in order to recover the complex scattering amplitude $S_{21}$. By scanning the position of the probe antenna in the measurement plane, we can measure the local electric field profile. We apply a deconvolution filter to account for the anisotropy of the probe and the finite aperture of the device. The complex fields are computationally back-propagated to the focal plane of the device. This analysis is repeated for a range of microwave frequencies within the $K_a$ band (26–40 GHz), and for both orthogonal polarizations of the input beam. To measure the scattering parameters for an orthogonal polarization, we rotate the device 90°. We also measure the cross-polarized fields by rotating the waveguide flange 90°.

## VII. Power flow analysis

We experimentally validate the design of volumetric scattering elements using a microwave analog by measuring the scattered fields. This analysis reveals close agreement between simulated and measured sorting efficiencies, roughly 40% across the measurement spectrum. However, the performance is lower than previously designed devices (Fig. 1 and 2). Here we account for the remaining energy that is lost in the system. From full-wave simulations we can quantify four sources of loss: reflection from the device, oblique scattering away from the focal plane, and absorption within the material (Fig. S7a). Both reflection and oblique scattering contribute equally to the overall loss in the system, independent of wavelength (Fig. S7b). We believe that these losses can be significantly mitigated by explicitly including them in the optimization algorithm, or through the appropriate choice of boundary conditions and anti-reflection coatings. Due to the agreement between simulated and experimental efficiencies, we conclude that absorption plays a negligible role, which is consistent with the extremely low extinction coefficient of polypropylene in the $K_a$ band [7].

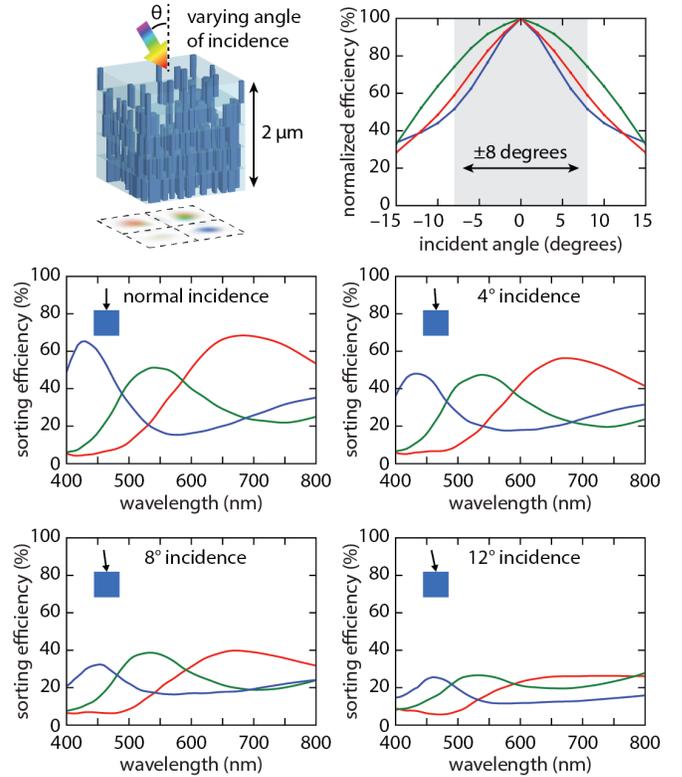

**Fig. S5.** Sorting performance at non-normal incidence. The layered device is optimized to sort light incident from above. The sorting performance degrades as the incident angle increases. Compared to the design at normal incidence, the efficiency is reduced by half at angles beyond ±8°.

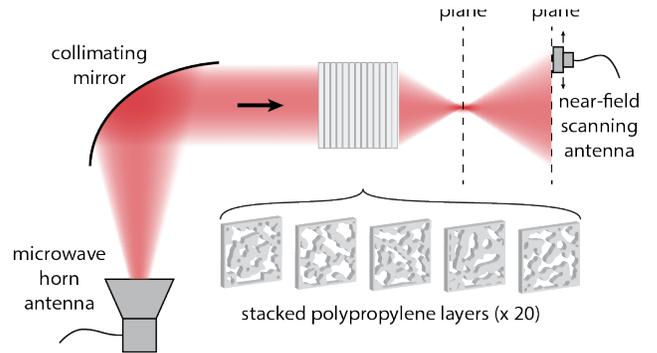

**Fig. S6.** Experimental characterization at microwave frequencies. Microwave device consists of 20 patterned polypropylene sheets, assembled into a cube. The cube is illuminated by a collimated microwave source operating in the $K_a$ band (26.5 – 40 GHz). Local electric fields are collected by a scanning near-field probe, and back-propagated to the focal plane.

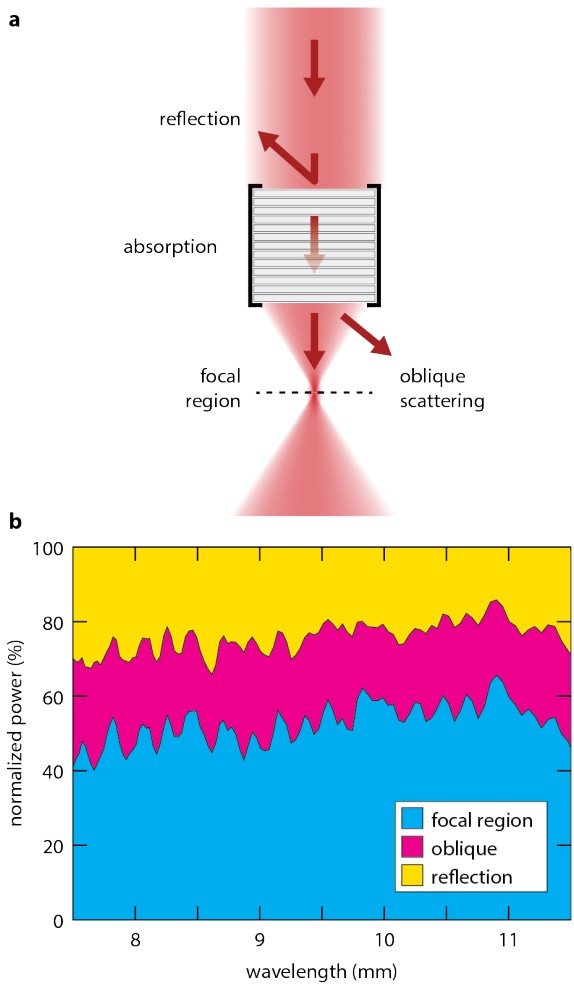

**Fig. S7.** Power flow analysis of microwave devices. (a) The incident microwave beam is deflected into different regions, contributing to lower overall sorting efficiency. (b) Distribution of optical power in simulations, including transmission, reflection, and oblique scattering. The power is normalized to the input aperture of the device.